\documentclass[english,12pt]{article}
\usepackage[utf8]{inputenc}
\usepackage[english]{babel}
\usepackage{graphicx}
\usepackage{slashed}
\usepackage{hyperref} 
\usepackage{geometry}
\usepackage{amsmath}
\usepackage{amssymb}
\usepackage{amsfonts}
\usepackage{multicol}
\usepackage{float}
\usepackage{caption}
\usepackage{subcaption}
\usepackage{cite}
\usepackage{amsthm}
\usepackage{xcolor}
\title{\bf Implications of Quantum Gravity for Dark Matter}
\author{Xavier Calmet\thanks{E-mail: X.Calmet@sussex.ac.uk; corresponding author.}$\ $ 
and Folkert~Kuipers\thanks{E-mail: F.Kuipers@sussex.ac.uk} 
\\
\\
{\em Department of Physics and Astronomy, University of Sussex,}\\ 
{\em Brighton, BN1 9QH, United Kingdom}
}

\begin{document}
\maketitle
%\vspace{2cm}
%
\begin{abstract}
In this essay we show that quantum gravity and the spin-statistics theorem have very interesting consequences for dark matter candidates. Quantum gravity can lead to fifth force type interactions that lead to a lower bound on the masses of bosonic candidates. In the case of fermions, the spin-statistics theorem leads to a lower bound on fermion masses. For both bosonic and fermionic dark matter candidates, quantum gravity leads to a decay of dark matter particles. A comparison of their lifetime with the age of the universe leads to an upper bound on their masses. For singlet scalar dark matter fields, we find $10^{-3}{\rm eV} \lesssim m\lesssim 10^{7}{\rm eV}$.
\end{abstract}

Essay written for the Gravity Research Foundation 2021 Awards for Essays on Gravitation.
\flushbottom
%\raggedbottom
%
\thispagestyle{empty}
\pagebreak
\pagenumbering{arabic}
 \newpage 
 \newpage

 In this essay we make a very simple point. Quantum gravity will generically generate portals between any hidden sector, including a dark matter sector in particular, and the visible sector of the universe described by the Standard Model of particle physics. The specific nature of these portals depends on the spins of the hidden sector particles, the symmetries of the particle physics sectors and those of quantum gravity. In the case of a hidden sector consisting of singlet boson dark matter particles, the consequences can be quite dramatic as quantum gravity generates operators mediating fifth force type interactions within the visible sector. There are however tight bounds from the E\"ot-Wash pendulum experiment \cite{Kapner:2006si,Hoyle:2004cw,Adelberger:2006dh,Leefer:2016xfu,Braginsky1972,Smith:1999cr,Schlamminger:2007ht,Adelberger:2009zz,Zhou:2015pna} on long-range new forces which leads to a lower bound on the mass of the dark matter particle. The spin-statistics theorem leads to a similar bound for light fermion dark matter particles \cite{Tremaine:1979we,DiPaolo:2017geq,Savchenko:2019qnn}. In addition, quantum gravity will generate operators that lead to a decay of the dark matter particle. The requirement that the dark matter decays slowly enough to still be present in today's universe leads to an upper bound on the masses of dark matter particles. 
 
 To illustrate our point, we shall focus on singlet scalar fields and singlet Dirac fermions, but our results can be trivially generalized to any spin and we shall simply quote the results for other spins. It is easy to convince oneself that perturbative quantum gravity only leads to operators of dimensions 8 and higher suppressed by at least four powers of the reduced Planck mass $M_{P}=2.435\times 10^{18}$ GeV. These operators are thus irrelevant for low energy experiments or observations. On the other hand, non-perturbative quantum gravity effects such as quantum black holes or gravitational instantons can lead to operators of dimension four and higher which have the potential of being much more important. 
 
  Everything couples to gravity as it is universal. Therefore, one can deduce that non-perturbative quantum gravity effects should generate effective operators between the hidden sector and the visible sector. These operators must be suppressed by the scale of quantum gravity as such interactions must vanish in the limit where $M_P \to \infty$, i.e. when gravity decouples. This implies that for operators of dimension four, the suppression must be exponential $\exp{(-M_P/\mu)}$ where $\mu$ is some infra-red scale. Furthermore, for dimension five and higher, the suppression factor is much weaker, namely $M_P^{-n+4}$ where $n\ge5$ is the dimension of the operator. Classifying these operators is easy, they can involve any of the fields of the Standard Model and any of the fields of the hidden sector. They must respect gauge symmetries and diffeomorphism invariance of the models involved in our theory, but could violate any global or discrete symmetry as long as they are not the reminiscence of a spontaneously broken gauge symmetry. Space-time symmetries such as Lorentz invariance are generically expected to be preserved albeit there are some string theory vacua known to break Lorentz invariance spontaneously \cite{Kostelecky:1988zi}.  Discrete symmetries space-time $C$, $P$ and $T$ symmetries  could be individually violated by quantum gravity, but the combination $CPT$ must hold.

Let us first consider a singlet scalar field from the hidden sector. Using the guiding principles highlighted above, we consider local operators that are generated by non-perturbative quantum gravity effects (see e.g. \cite{Calmet:2019jyz,Calmet:2019frv,Calmet:2020rpx,Calmet:2020pub,Holman:1992us,Barr:1992qq,Calmet:2014lga,Giddings:1988cx,Coleman:1988tj,Gilbert:1989nq}):
\begin{equation}
	O_1 = \frac{c_\phi}{M_{\rm P}}  \, \phi \, F_{\mu\nu}F^{\mu\nu},
\end{equation}
where $M_{\rm P}=2.4 \times 10^{18}$ GeV is the reduced Planck scale, $\phi$ is the scalar dark matter field, and $F_{\mu\nu}$ is the electromagnetic field tensor of the Standard Model.  The Wilson coefficient $c_\phi$ is expected to be of order unity. This operator is an example of a fifth force type interaction mediated by the hidden sector field $\phi$ between the particles of the Standard Model. The results from the E\"{o}t-Wash torsion pendulum experiment that searches for fifth forces imply that $m_\phi \gtrsim 10^{-3}\, {\rm eV}$ \cite{Calmet:2019jyz,Calmet:2019frv,Calmet:2020rpx}.  

The same operator induces the decay of the dark matter scalar \cite{Calmet:2009uz,Mambrini:2015sia} with a decay width $\Gamma \sim  m_\phi^3/(4 \pi M_P^2)$. This leads to an upper bound $m_\phi \lesssim 10^{7}{\rm eV}$ from the requirement that the dark matter candidate lives long enough to still be present in today's universe.  Quantum gravity thus enables to restrict the mass of any singlet scalar particle to be in the range:
\begin{equation}
10^{-3}{\rm eV} \lesssim m_\phi \lesssim 10^{7}{\rm eV},
\end{equation}
independently of its potential non-gravitational couplings to Standard Model particles or self-interactions. Note that these bounds would not apply to a gauged dark matter scalar field as only dimension six operators would be generated by quantum gravity. In that case, one has $m_\phi \gtrsim 10^{-22}{\rm eV}$  \cite{Calmet:2020rpx}, and the upper bound disappears.

For spin $1/2$ fermions $\psi$, quantum gravity leads to an upper bound on the mass of the dark matter candidate \cite{Calmet:2009uz,Calmet:2020pub,Mambrini:2015sia,Boucenna:2012rc}  as it could decay to the Standard Model fields, while a lower bound comes from the Pauli exclusion principle. We consider the operator \cite{Calmet:2009uz,Calmet:2020pub,Mambrini:2015sia}:
\begin{equation}
	O_\psi = \frac{c_\psi}{M_{\rm P}}  \, \bar \psi \tilde H^\dagger \slashed{D} L,
\end{equation}
where $H$ is the Higgs doublet of the Standard Model with $\tilde H= -i \sigma_2 H^\ast$. This operator implies that the singlet right-handed fermion $\psi$ can decay to an off-shell $Z$ boson and a neutrino, the $Z$ boson then decays to two light fermions. Requiring that the fermion singlet lives long enough to still be present today imposes an upper bound on its mass. One finds $m_\psi < 10^{10}{\rm eV}$  using $\Gamma=v^2 G_F^2 m_\psi^5/(192 \pi^3 M_P^2)$ where $G_F$ is the Fermi constant and $v=246$ GeV the electroweak vacuum expectation value.

Because of the spin-statistics theorem, only a limited amount of fermions can be present in a galaxy with momenta below the escape velocity. Together with the assumption that the fermions must account for the observed dark matter density in a typical galaxy this leads to a lower bound on the mass of the fermions \cite{Tremaine:1979we}. The bounds on the mass of the dark fermion are then given by 
\begin{equation}
	10^{2}{\rm eV} \lesssim m_\psi \lesssim 10^{10}{\rm eV}.
\end{equation}

We can easily generalize our results to other spins. The mass of a pseudo-scalar dark matter candidate, an axion like particle, $a$ if quantum gravity violates parity (and time reversal invariance) is bounded by:
\begin{equation}
10^{-3}{\rm eV} \lesssim m_a \lesssim 10^{7}{\rm eV}.
\end{equation}
On the other hand, if quantum gravity preserves parity,  we find \cite{Calmet:2020rpx,Calmet:2020pub,Mambrini:2015sia}
\begin{equation}
 10^{-21}{\rm eV} \lesssim m_a \lesssim 10^{7}{\rm eV}.
 \end{equation}
For a vector field $V$, the first quantum gravitational operator that can be generated non-perturbatively is of dimension 6 because of the chiral nature of fermions in the Standard Model. We find
\begin{equation}
	10^{-22}{\rm eV} \lesssim m_V \lesssim 10^{7}{\rm eV}.
\end{equation}
For a massive spin-2 dark matter field, similar considerations lead to  \cite{Calmet:2019frv,Calmet:2020pub}
\begin{equation}
10^{-3}{\rm eV} \lesssim m_2 \lesssim 10^{7}{\rm eV}.
\end{equation}

Our bounds are derived assuming the worst case scenario for quantum gravity, namely that it has only one scale and that this scale is the traditional reduced Planck scale i.e. $2.4 \times 10^{18}$ GeV. In other words, we assumed that quantum gravity is as weak as possible. Our bounds become much more stringent if the effective scale of quantum gravity is below the traditional Planck scale as it is the case in models with large extra-dimensions \cite{ArkaniHamed:1998rs,Randall:1999ee} or a large number of fields \cite{Calmet:2008tn} where it could be in the TeV region. 

Finally let us emphasize that a discovery of a dark matter particle with a mass violating our bounds could indicate that there is a new gauged force in the hidden sector. Indeed, a violation of these bounds would imply that the dimension 5 operators considered here are not generated. For example, a gauged scalar field $\Phi$, would not lead to a dimension 5 portal operator, but rather  a dimension 6 one. In that case, pairs of dark matter particles could annihilate to standard model particles, but dark matter would be stable and our higher bound on their masses would not apply. A dimension 6 operator $\Phi^\dagger \Phi F_{\mu\nu} F^{\mu\nu}$ would lead to a fifth force as well, but the bounds on the Wilson coefficients of such operators and thus on the masses of these scalar fields, are much weaker than those on dimension 5 operators. The discovery of a dark matter particle with a mass violating our bounds would thus imply some much richer structure in the hidden sector, for example, in the form of new dark forces.

\section*{Acknowledgments}
 The work of X.C.~is supported in part  by the Science and Technology Facilities Council (grants numbers ST/T00102X/1, ST/T006048/1 and ST/S002227/1). The work of F.K.~is supported by a doctoral studentship of the Science and Technology Facilities Council.

\end{document}